# Efficient condensation on spiked surfaces with superhydrophobic and superhydrophilic coatings

Kai-Xin Hu[a,b,1], Yin-Jiang Chen[a,b], Bin-Wei Tang[a,b], Yue-Qun Tao[c,d], Qiu-Sheng Liu[c,d], Qi-Sheng Chen[c,d]

[a]Zhejiang Provincial Engineering Research Center for the Safety of Pressure Vessel and Pipeline, Ningbo University, Ningbo, Zhejiang 315211, China

[b]Key Laboratory of Impact and Safety Engineering (Ningbo University), Ministry of Education, Ningbo, Zhejiang 315211, China

[c]School of Engineering Science, University of Chinese Academy of Sciences, Beijing 100190, China

[d]Key Laboratory of Microgravity, Institute of Mechanics, Chinese Academy of Sciences, Beijing 100190, China

**Abstract**

Steam condensation on the surface of a solid is a widely observed mode of energy transfer in nature and various industrial applications. The condensation efficiency is closely related to the material properties and geometric morphology of the solid surface, as well as the method of liquid removal. Despite many surface modification strategies at the micro and nano levels having been employed to enhance steam condensation, understanding how to regulate steam condensation and liquid removal on complex surface morphologies remains incomplete. Here, we report a method that uses superhydrophilic and superhydrophobic coatings as well as spiked surfaces to achieve efficient steam condensation and rapid removal of the liquid. We reveal that on a copper plate with millimeter-scale spikes, hydrophobic spiked surfaces facilitate the dropwise condensation, while hydrophilic bottom grooves promote liquid spreading, and the suction in capillary gaps can promptly remove the condensate liquid. This method achieves a high condensation efficiency without relying on gravity. Additionally, we demonstrate that the condensation on spiked surfaces has a significant suction effect, as it continuously attracts the steam flow, thereby altering the flow direction. This finding provides a new approach for efficient steam condensation technology and opens up a promising pathway for providing circulation power and improving condensation efficiency in phase-change heat transfer devices such as heat pipes.

[1]Corresponding author, Email: hukaixin@nbu.edu.cn

Steam condensing into liquid is a common natural phenomenon. This phase transition process releases a large amount of latent heat, making it an efficient method of energy transfer. It plays an increasingly important role in industries such as power generation[1], refrigeration[2], water collection[3,4], electronic device cooling[5], and aerospace thermal management[6]. When steam condenses on a solid surface, if the liquid can wet the surface well, it can spread into a film on the wall, known as film condensation. When the liquid has poor wettability on the solid wall and condenses into small droplets, it is called dropwise condensation. The latter, due to its heat transfer performance being one to two orders of magnitude higher than the former, has become a goal pursued by researchers for many years[7-9]. However, dropwise condensation is not easy to maintain. Even if the surface properties such as substrate roughness and wettability remain unchanged during the condensation process, as the condensation droplets increase, they can easily merge and transform into film condensation[10]. Timely removal of the condensate is the key to improving condensation efficiency[11-13].

Recently, a significant amount of groundbreaking research has confirmed that the combination of hydrophilic and hydrophobic surfaces can achieve dropwise condensation, enhancing heat transfer efficiency. The hydrophilic areas facilitate droplet nucleation, while the hydrophobic surfaces aid in droplet detachment. Combining these advantages can significantly improve condensation efficiency[14-16]. Additionally, droplets can spontaneously migrate from the weakly wettable hydrophobic regions to the strongly wettable hydrophilic regions, which aids in the removal of condensate droplets[17]. Although excellent work has been done to achieve sustained dropwise condensation to some extent[18-22], how to integrate complex surface morphology[23,24] and micro-nano level surface modification techniques[25,26] to achieve rapid and efficient steam condensation and liquid migration is still in its infancy.

Here, we report a new discovery that spiked surfaces combined with superhydrophilic/superhydrophobic coatings can efficiently condense steam into liquid and promptly remove it. Taking a horizontal plate with spikes as an example, the superhydrophobic coating on the spiked surfaces facilitates the dropwise condensation, while the superhydrophilic coating on the bottom grooves ensures uniform distribution of condensate. The glass rods in contact with the edge of the condensation zone can utilize capillary forces to promptly absorb the condensate, preventing flooding on the condensation zone. Thus, we can achieve a high condensation efficiency without relying on gravity, where the condensate flow rate surpasses that of traditional vertical plates and the heat flux density reaches 973.8kW/$m^2$ . By observing the condensation process, adjusting the wettability patterns, and comparing with vertical substrates, we analyze the fundamental mechanisms of enhanced condensation. By using the spiked plate and coatings, we fabricate a heat pipe with a power of 480W. More importantly, we demonstrate that the condensation on the spiked surfaces exhibits a significant suction force for high-temperature steam, which can promote directional movement of the steam flow, potentially providing a new driving force for circulation in phase-change heat transfer devices.

## Results

**The condensation on spiked surfaces**. Figure 1a illustrates our experimental system, which comprises a steam generation device, a condensation device, a collection device, and a cooling device. The high-temperature steam (100.4°C) generated by heating the water in the flask flows through the silicone tube and is ejected from the nozzle. It then condenses into liquid on the condensation area of the copper plate, which includes spiked surfaces and microgrooves at the bottom, and accumulates into a liquid film. As the condensate increases, the boundary of the liquid film continuously expands until it comes into contacts with the glass rods in the collection area. The capillary force generated by the gaps between the glass rods and the copper surface can draw the condensed liquid to the other end of the copper plate. Then, under the action of gravity, the liquid drips into the beaker, thereby achieving the collection of the condensate. The heat released by the condensation is carried away by the cooling circuit, which includes a cooling copper block below the condensation zone and a constant-temperature water tank that provides cooling water for it. Supplementary Movie 1 shows the working status of the system.

To enhance the steam condensation and liquid migration effects, we apply coatings with different wettabilities to the surfaces of spikes and grooves in the condensation area, whose three-dimensional structure is displayed in Figure 1b. The superhydrophilic coatings are used in the surfaces of the copper plate and glass rods in the collection area. Under the same steam conditions and ambient temperature (24°C), we study the condensation effects of copper plates with different spike heights (Figure 1c). The results indicate that among various combinations of coatings, the optimal condensation effect is achieved when the spikes and grooves are coated with superhydrophobic and superhydrophilic coatings, respectively. Among three types of spikes, the copper plate with a spike height $h=$ 5mm exhibits the best condensation effect.

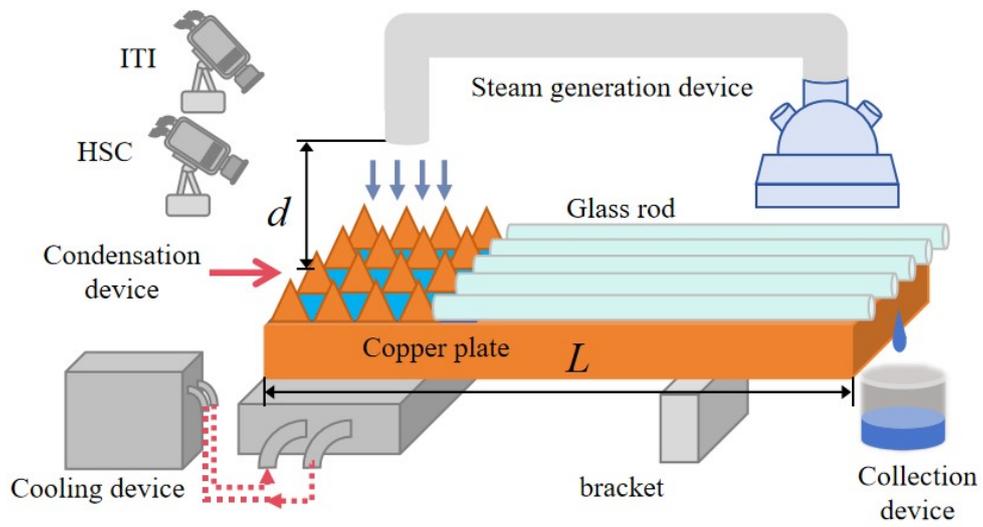

a

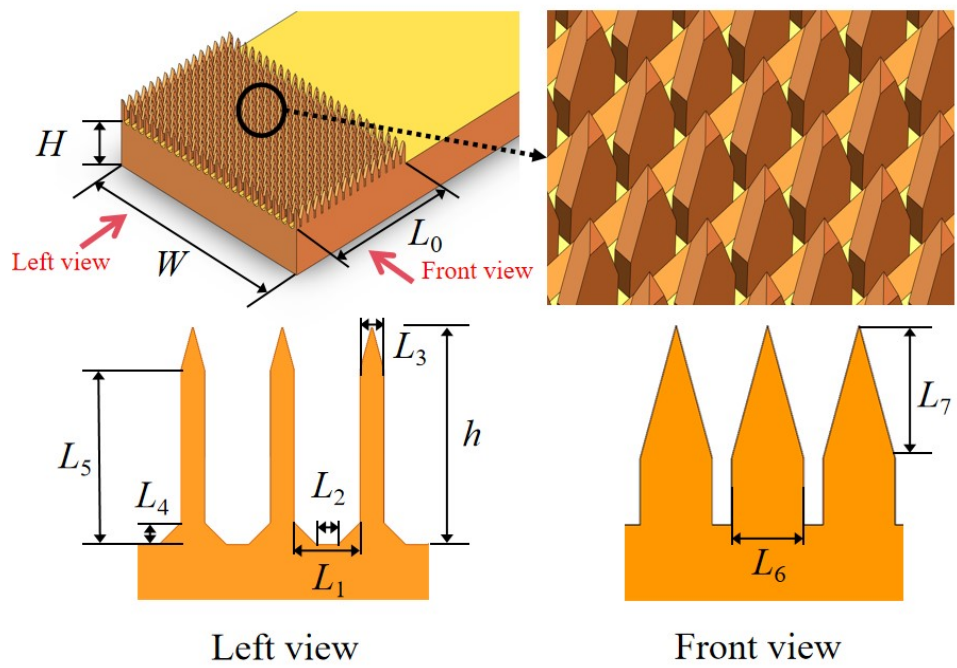

b

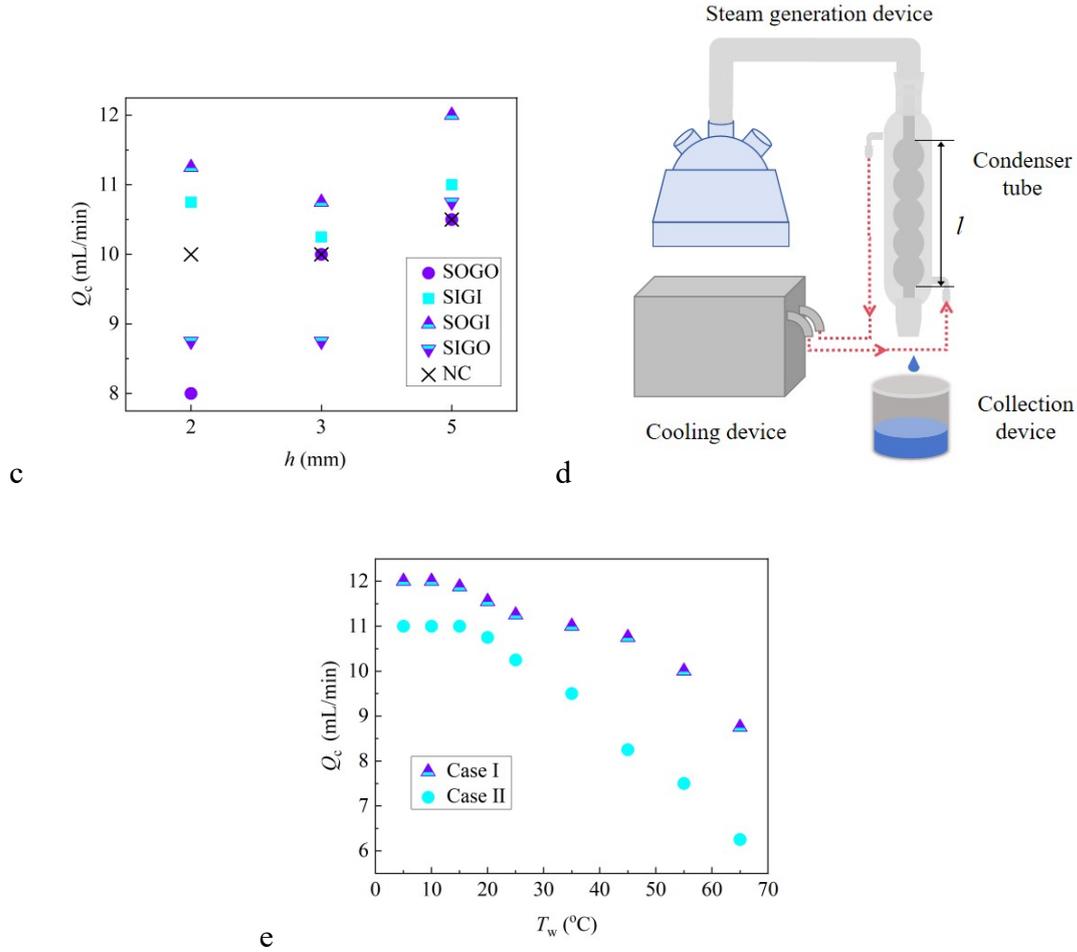

Figure 1 Steam condensation on a copper plate with spikes. **a**, Experimental system composed of a steam generation device, a condensation device, a collection device, and a cooling device. A high-speed camera (HSC) and an infrared thermal imager (ITI) are used to record the condensation process and the temperature distribution on the solid surface, respectively. The outlet of steam pipe is positioned at a distance of $d$=5mm from the spikes. The total dimensions of the copper plate are $L$=150mm, $W$=50mm, $H$=10 mm. **b**, Schematics of the three-dimensional structure of the condensation zone, which includes spikes and microgrooves. Three different copper plates with varying spike heights ($h$=2mm, 3mm and 5mm) are used. $L_0$=30mm, other parameters ($L_1$~$L_7$) are detailed in Table 2. **c**, The volumetric flow rate of condensate $Q_c$ under different conditions. SOGI represents a superhydrophobic coating on the spikes and a superhydrophilic coating on the grooves; SIGO represents a superhydrophilic coating on the spikes and a superhydrophobic coating on the grooves; SIGI represents superhydrophilic coatings on both spikes and grooves; SOGO represents superhydrophobic coatings on both spikes and grooves. NC represents no coating. **d**, Steam condensation using spherical condenser tubes, with three different lengths $l$=200mm, 300mm, and 400mm. **e**, The variation of $Q_c$ with the temperature of cooling water $T_w$. Case I stands for the SOGI coatings with $h$=5mm, Case II stands for the SIGI coatings with $h$=2mm.

When a superhydrophilic coating is applied to the surface of the microgroove, the condensate in the center of the condensation zone will continuously spread under the capillary force of microgrooves until it reaches the boundary of the condensation zone, and then flows along the glass rod towards the collection zone. It can be seen that the wettability of the microgroove surface is very conducive to the timely drainage of condensate. Dropwise condensation and filmwise condensation will occur on the spiked surface when it is coated with superhydrophobic and superhydrophilic coatings respectively. The former shows a higher condensation efficiency. Therefore, the condensation efficiency of SOGI (a superhydrophobic coating for the spike and a superhydrophilic coating for the microgroove) is the best, followed by SIGI (Both the spikes and the microgroove surfaces are coated with a superhydrophilic coating).

When a superhydrophobic coating is used for the microgroove, the condensate does not easily spread, leading to accumulation in the condensation zone and causing flooding. The liquid film continuously grows in size and thickness until it makes contact with the glass rod. Although the drainage of the glass rod can lower the water level in the condensation zone, the thickness of the liquid film remains high when the system reaches stability, so the condensation efficiency is sometimes worse than that of no coating. The coating on the spiked surface also affects the condensation, but due to the hydrophobic nature of microgrooves at the bottom, the condensate cannot be drained in a timely manner, resulting in an overall poor condensation efficiency.

Apart from the surface wettability, the spike height also has a significant impact on the condensation efficiency. $Q_c$ at $h$=5mm is larger than other two cases for different coating combinations, while the variation of $Q_c$ with $h$ is not monotonous for some coatings.

In order to measure the steam flow rate, we connect the steam outlet to a spherical condenser tube and measure the condensate flow rate using a burette (as shown in Figure 1d). Three different lengths of condenser tubes are applied in the measurement. The results suggest that the condensate flow rates from all three condenser tubes are the same $Q_0 = 12.75 \text{mL/min}$, with a liquid temperature of $T_0 = 25°C$ and density of $\rho_0 = 0.997 \times 10^3 \text{kg/m}^3$. Therefore, it can be assumed that all the steam has condensed into liquid. Since the maximum value of condensate flow rate in Figure 1c is $\max(Q_c) = 12 \text{mL/min}$, the optimal condensation efficiency is $\eta = 12/12.75 \approx 94.1\%$. At a temperature of 100.4°C, the saturated steam density is $\rho = 0.606 \text{kg/m}^3$, thus the steam flow rate is $Q = Q_0 \cdot \rho_0/\rho = 349.6 \text{cm}^3/\text{s}$. As the inner diameter of the nozzle $r = 0.95 \text{cm}$, we can determine that the average steam velocity at the nozzle is $v = Q/(\pi r^2) = 1.233 \text{m/s}$.

Figure 1e shows the variation of the condensate flow rate $Q_c$ with the temperature of cooling water $T_w$. It can be found that as $T_w$ increases, $Q_c$ remains unchanged in the early stage; when $T_w$ exceeds a certain value, $Q_c$ gradually decreases. Therefore, in other experiments of this paper, we maintain $T_w$=5°C to maximize the condensation.

**Properties of the condensation area**. In order to clarify the condensation process in a steady state, we observe the condensation area using a high-speed camera and an infrared thermal imager.

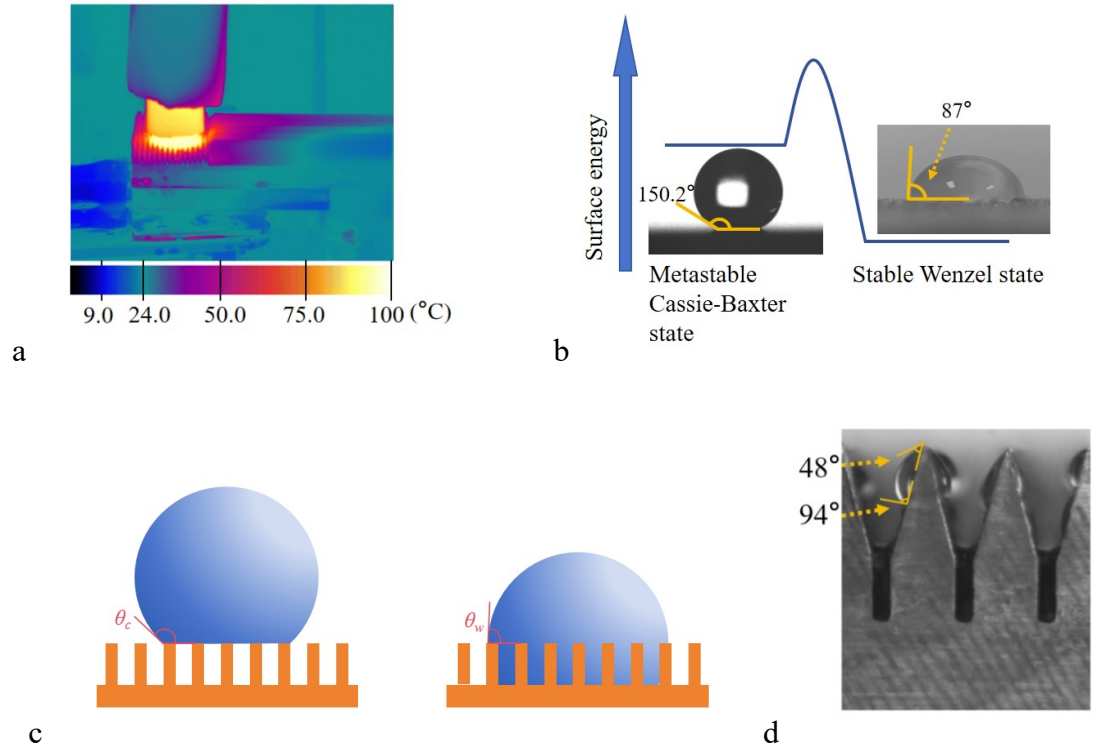

Figure 2 Properties of the condensation area in a horizontal copper plate with a spiked surface. **a**, The temperature distribution of a spiked surface as captured by the infrared thermal imager. The spike height is $h$=5mm, and the SOGI coating is used. The heat flux density within the high-temperature circular area reaches $973.8 \text{kW/m}^2$. **b**, Droplets located on a superhydrophobic surface in two different configurations. The droplet dropped onto the surface at room temperature (24°C) has a contact angle of $\theta_c = 150.2°$ (Cassie-Baxter state), while the droplet formed by condensation of steam on the surface has a contact angle of $\theta_w = 87°$ (Wenzel state). The former has a higher surface energy. **c**, Schematics depicting a droplet in contact with a rough surface, illustrating Cassie-Baxter contact on the left and Wenzel contact on the right. **d**, The contact angles of a droplet formed by condensation on the spiked surface in a horizontal copper plate. The spiked surface is coated with a superhydrophobic coating. The advancing angle of the droplet is 94°, while the receding angle is 48°.

Figure 2a shows the temperature distribution of the condensation area. It can be observed that the high temperature area on the spiked surface is mainly concentrated within a circular area with a radius of $r_1$=1.25cm. However, due to water's strong absorption of infrared radiation, the temperature values in the image are not accurate. The temperatures of liquid in the condensation area is measured by a digital thermometer to be 100.4°C. Now we calculate the heat flux density within the high-temperature circular area in the following.

As the volume flow rate of steam is $Q = 349.6 \text{cm}^3/\text{s}$, the mass flow rate is $Q_m = \rho Q = 2.12 \times 10^{-4} \text{kg/s}$. The latent heat of vaporization is $\Delta H = 2255.87 \text{kJ/kg}$, so the heat flux generated by condensation is $P_1 = Q_m \cdot \Delta H = 0.478 \text{kW}$, and the heat flux density is $q = P_1/(\pi r_1^2) = 973.8 \text{kW/m}^2$, which is much larger than the value $655 \text{kW/m}^2$ in a 3D hybrid surface consisted of superhydrophobic nanowire arrays and hydrophilic microchannels[27].

When a superhydrophobic coating is used in a smooth horizontal copper plate, the contact angle formed when a liquid droplet is dropped onto the surface at room temperature (24°C) is 150.2°. On the contrary, when the droplet is formed by the condensation of steam on a horizontal surface, the contact angle is only 87° (Figure 2b). However, the superhydrophobic coating is not damaged. If the copper plate is heated and dried, then dropping the droplet on the surface at room temperature, the contact angle can be restored to 149.5°.

The above phenomenon indicates that droplets can be located on a superhydrophobic surface in two different configurations. When a droplet is able to completely wet every area of a rough surface, it is in the Wenzel state[28] and satisfies the following equation $\cos\theta_Y = \cos\theta_w / r_m$, where $\theta_Y$ represents the Young's equilibrium contact angle, $\theta_w$ is the apparent contact angle of the droplet on the rough surface, and $r_m$ is the roughness measure of the solid surface, with its specific expression being $r_m = \frac{1}{A}\iint_A \sqrt{\left[1+\left(\frac{\partial z(x,y)}{\partial x}\right)^2+\left(\frac{\partial z(x,y)}{\partial y}\right)^2\right]}dxdy$. Here, $A$ is the cross-sectional area of the observed region, and the height $z$ is a function of the horizontal coordinates $x$ and $y$.

We use a contact angle measuring instrument and a laser scanning confocal microscope to observe the apparent contact angle and the microstructure of the coating, respectively, finding $\theta_w = 87°$, $r_m = 1.2896$, and $\cos\theta_Y = 0.0406$.

In another scenario, the droplet is suspended at the tips of a rough surface, at which point the droplet is in the Cassie-Baxter state[29]. The apparent contact angle in this state $\theta_c$ satisfies the following relationship $\cos\theta_c = -1 + \phi(\cos\theta_Y + 1)$, where $\phi$ is the fraction of the drop footprint that is in contact with the solid surface. In the experiment, $\theta_c = 150°$, and thus, we are able to determine $\phi = 0.1287$.

Both the Cassie-Baxter and Wenzel states can be stable, and droplets will adopt the thermodynamically more stable state. The Wenzel state is preferred if $\cos\theta_Y > (\phi-1)/(r-\phi)$, and the Cassie-Baxter state is preferred otherwise (Khandekar, 2020). Substituting the relevant data, we find that this inequality is satisfied, so the Cassie-Baxter and Wenzel states correspond to local and global energy minima, respectively, and are referred to as metastable and stable states, respectively. The droplets formed by condensation on superhydrophobic coatings have smaller contact angles, which are in the Wenzel state. This phenomenon occurs because the steam can condense in the depressions of the rough surface, which allows the condensed liquid to come into full contact with the surface, minimizing the likelihood of air pockets forming. From the perspective of condensation, the Wenzel state has a larger contact area than the Cassie-Baxter state, leading to a higher heat flux and thus promoting droplet growth. However, from the perspective of droplet migration, the former has a smaller contact angle, resulting in a higher migration resistance.

For a horizontal copper plate with a spiked surface using SOGI coatings, the droplets formed by condensation on the spiked surface exhibit a significant contact angle hysteresis. Figure 2d shows that the advancing angle of one droplet is 94°, while the receding angle is 48°. This indicates that gravity has a great impact on the morphology of the sessile droplets.

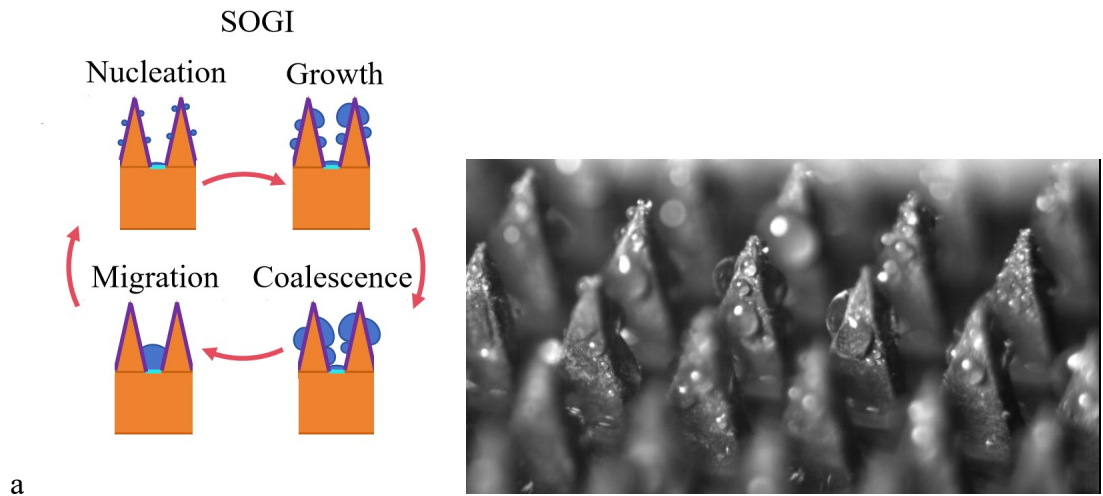

a

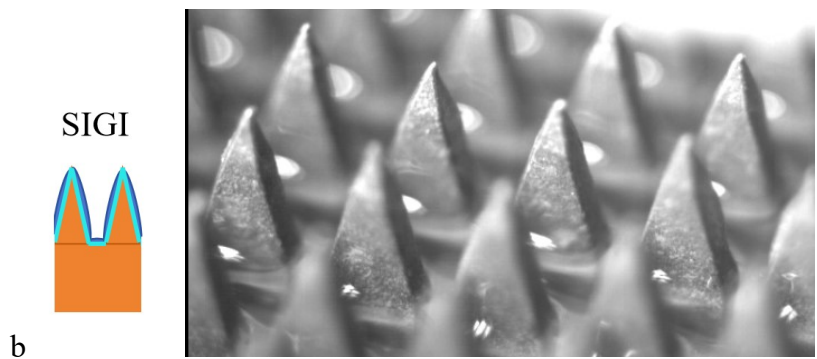

b

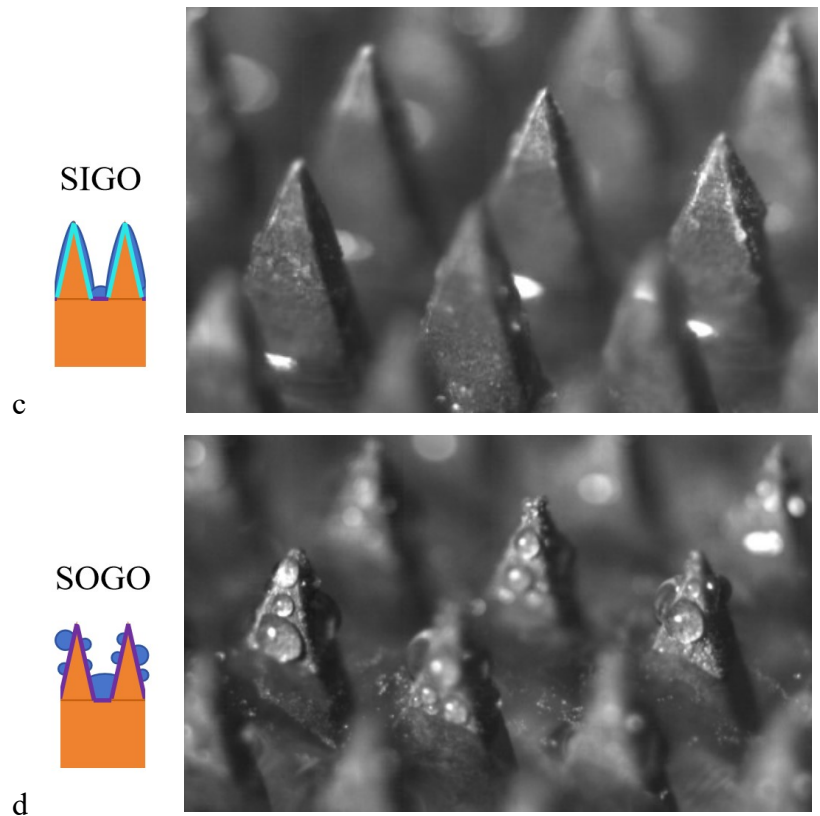

Figure 3 The condensation process in a horizontal copper plate with a spiked surface. **a**, When using the SOGI coatings, steam undergoes a dropwise condensation on the spiked surface, repeating a cycle of four stages: nucleation, growth, coalescence, and migration. **b**, When using the SIGI coatings, a film condensation occurs on the surfaces of spikes and microgrooves. **c**, When using the the SIGO coatings, there is a thin liquid film on the spiked surface, while liquid easily accumulates at the bottom. **d**, When using the SOGO coatings, a dropwise condensation occurs on the spiked surfaces, while a thick liquid film forms at the bottom.

Figure 3 shows the condensation process under various coatings. For the case of optimal condensation efficiency (SOGI), droplets nucleate and continuously grow on the spiked surface without migration (Figure 3a and Supplementary Movie 2). As the droplets grow to a critical size (average diameter $D_1$=1.18mm, maximum diameter $D_2$=1.32mm), multiple droplets coalesce and come into contact with the liquid film on the superhydrophilic coating. Due to the surface tension, the liquid quickly migrates from the spiked surface to the microgrooves, and then new droplets nucleate on the spiked surface, starting the next cycle. In the next Section, we examine the influence of gravity on the condensation process, finding that gravity does not have a substantial impact on the mechanism of dropwise condensation and droplet migration described above.

For the case of suboptimal condensation efficiency (SIGI), steam undergoes a filmwise condensation on the surfaces of spikes and microgrooves (Figure 3b and Supplementary Movie 3). The hydrophilicity of the surface ensures that the condensed liquid is evenly distributed in the condensation area, resulting in a very thin liquid film. For the SIGO coatings (Figure 3c and Supplementary Movie 4), a filmwise condensation occurs on the spiked surface, while the superhydrophobic nature of the bottom allows for easy accumulation of liquid. For the SOGO coatings (Figure 3d and Supplementary Movie 5), steam undergoes a filmwise condensation on the spiked surface, but the accumulation of liquid at the bottom results in a thick liquid film.

In summary, we can see that the optimal condensation performance is achieved when a dropwise condensation occurs on the solid surface, followed by rapid removal of the liquid through the surface tension after droplet coalescence. The coating combination that corresponds to this effect is SOGI. On the contrary, the bottom surface with hydrophobic properties has a strong water-holding capacity, leading to a poor condensation efficiency.

**Comparison with other cases**. Previous researchers often used vertical flat plates for condensation, utilizing gravity to drain away the liquid produced by the condensation of steam[30,31]. Additionally, applying an electric field to the steam can promote condensation to a certain extent[32-35]. To further illustrate the effects of gravity and electric fields, we also conduct condensation experiments by using inclined plates and electric fields.

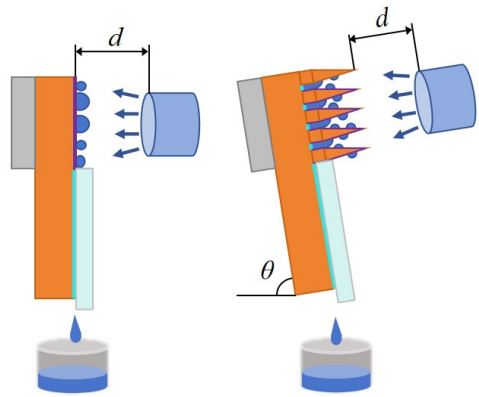

a

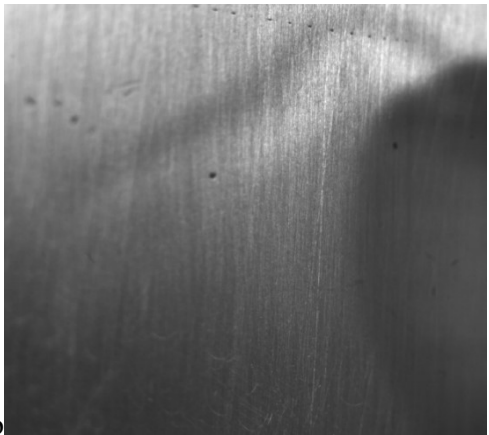

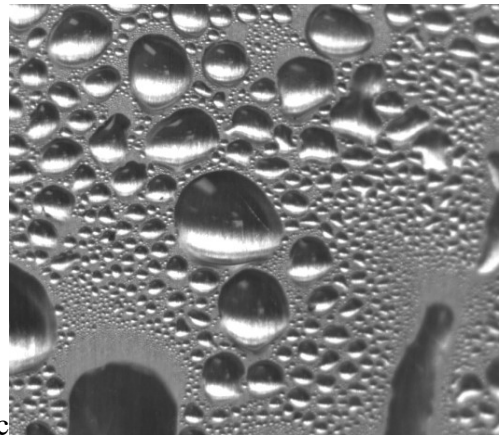

b c

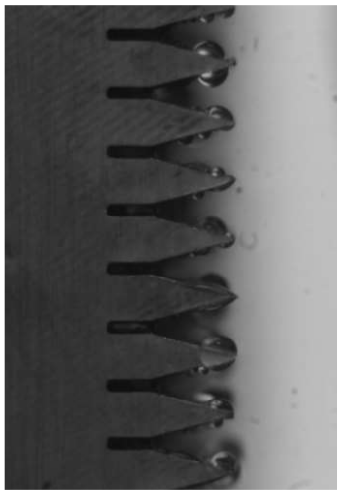

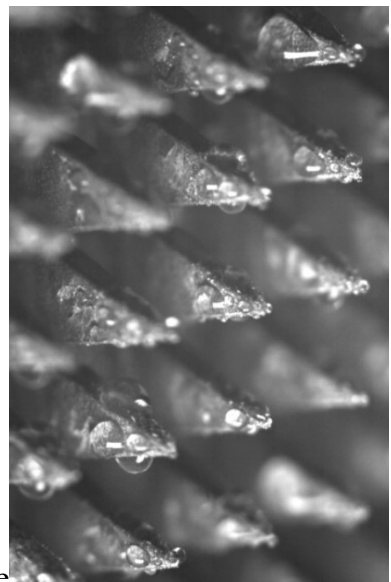

d e

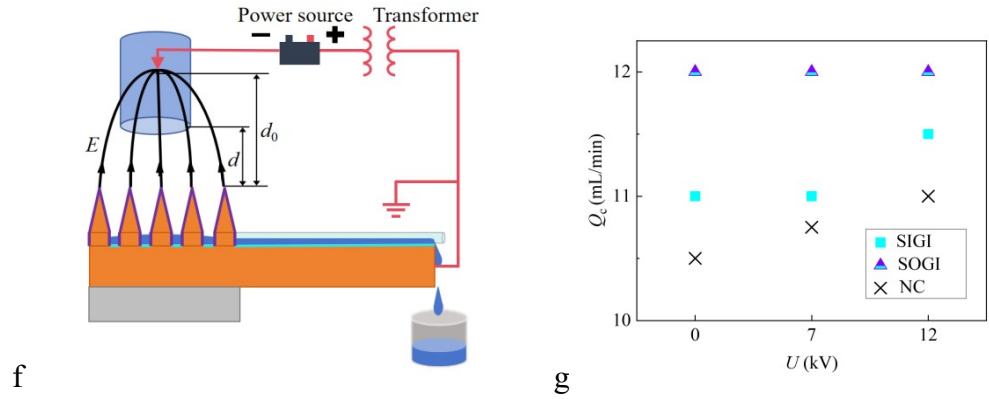

f  g

Figure 4  Condensation in other conditions. **a**, Schematics of steam condensation on the vertical smooth and inclined spiked ($h$=5mm) surfaces. The distance between the steam outlet and the spikes or smooth plate is $d$=5mm. **b** Filmwise condensation on the vertical smooth surface with a superhydrophilic coating. **c**, Dropwise condensation on the vertical smooth surface with a superhydrophobic coating. **d** & **e**, Front view and side view of dropwise condensation on spiked surfaces with SOGI coatings. **f**, Schematics of steam condensation on a horizontal plate in an electric field. The probe is connected to the negative electrode and placed at a distance $d_0$=20mm from the spikes, while the positive electrode is grounded. **g**, Condensation flow rate on the spiked surface ($h$=5mm) at different voltages and coatings.

Figure 4a is a sketch of our experiment on steam condensation using two types of surfaces, namely smooth and spiked ($h$=5mm) surfaces. Properties at different inclination angles and coatings are displayed in Table 1. The smooth surfaces are coated with either superhydrophilic or superhydrophobic coatings. In the vertical plate ($\theta$=90°), the solid surface experiences a dropwise condensation in Figure 4b and a filmwise condensation in Figure 4c. However, the maximum flow rate ($Q_c$ =11.5mL/min) is still lower than that of the horizontal plate ($\max(Q_c)$=12mL/min). For the horizontal smooth plate ($\theta$=0°), a liquid film appears on the solid surface. Due to the absence of gravity to promptly remove the liquid, the flow rate $Q_c$ becomes very low.

For the spiked surface using SOGI coatings, the direction of gravity has an impact on the morphology of sessile droplet (Figures 4d, 4e and and Supplementary Movie 6). In Table 1, as the inclination angle $\theta$ increases, the condensation rate slightly decreases, while the average cycle of dropwise condensation $T_c$ is sensitive to $\theta$. We can find that droplets always migrate from the spiked surface towards the microgrooves in these cases, and gravity influences the droplet morphology while exerting a small effect on the condensation flow rate. The surface tension of the liquid is the primary driving force that causes droplet migration. In addition, for the horizontal case ($\theta=0º$), the condensation flow rate of the spiked surface is much larger than that of the smooth surface.

**Table 1   Condensation Properties at different inclination angles and coatings**

| | Smooth Surface | | | | Spiked Surface | | | |
|---|---|---|---|---|---|---|---|---|
| | $\theta=90º$ | | $\theta=0º$ | | $h$=5mm, SOGI | | | |
| | SO | SI | SO | SI | $\theta=0º$ | $\theta=30º$ | $\theta=60º$ | $\theta=90º$ |
| $Q_c$ (mL/min) | 11.5 | 11.5 | 6.8 | 7.5 | 12.0 | 11.5 | 11.5 | 11.5 |
| $T_c$ (s) | 0.39±0.10 | -- | -- | -- | 0.91±0.10 | 0.94±0.10 | 0.88±0.28 | 1.60±0.80 |

SO and SI stand for the superhydrophilic or superhydrophobic coatings, respectively.

Figure 4f shows the schematic of our experimental setup for promoting condensation on a horizontal plate using a high-voltage electric field. By comparing the condensation flow rate under different voltages (Figure 4g), we find that the maximum of $Q_c$ does not change with the voltage, while in other cases, $Q_c$ increases with the voltage. This suggests that the high-voltage electric field has a certain promotional effect, but the condensation efficiency with the SOGI coatings has reached its limit and cannot be further enhanced by the electric field.

In Figure 1, we can find that the spiked surface with SOGI coating has the best condensation efficiency. It needs to be noted that this condensation mechanism does not depend on gravity. In the condensation zone, the driving force for liquid flow from hydrophobic spiked surfaces to microgrooves mainly comes from the surface tension, and in the collection zone, the driving force of liquid migration is the suction force of the capillary gap. Gravity is only used to make the liquid droplets fall into the beaker at the right end of the copper plate. If other forces (such as using a syringe to suction) remove the liquid at the right end, they can completely replace gravity, allowing the above steam condensation process to continue. In our experiment shown in Figure 5a and Supplementary Movie 7, a piece of copper foam is placed above a heating stage and connected to the right end of the copper plate. Thus, the liquid can flow into the copper foam and vaporize. In this way, the liquid can continuously flow from the condensation zone to the copper foam and evaporate, while the steam condensation at the left end can still continue.

If we redirect the steam from the right end to the left end for condensation and place the entire system in a sealed container (as shown in Figure 5b), the system becomes a heat pipe, continuously absorbing heat from vaporization at the heating section and releasing heat through condensation at the cooling section, while using capillary force to return the condensed liquid from the cooling section to the heating section. Figure 5c shows our experiment of heat pipe using the structure in Figure 5b. The heating block below the right end of the heat pipe is connected to the power supply to heat the pipe, while its left end is inserted into a water bath, which is cooled with ice water. We have set up a temperature measurement point on the outer wall surface of the heating, adiabatic and cooling sections, respectively. When the power of the heating block reaches 480W, the system can still operate stably, with the temperatures of each part being as follows: $T_h$=87°C (heating section), $T_a$=91°C (adiabatic section) and $T_c$=64°C (cooling section). Therefore, we see the spiked surface described in the present work can be easily applied to phase-change heat transfer devices, and the power of the heat pipe we presented in Figure 5c is really high at this size.

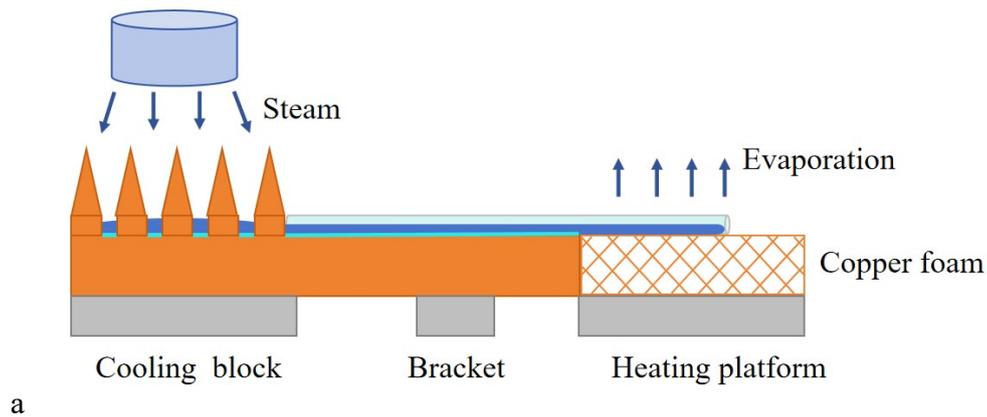

a

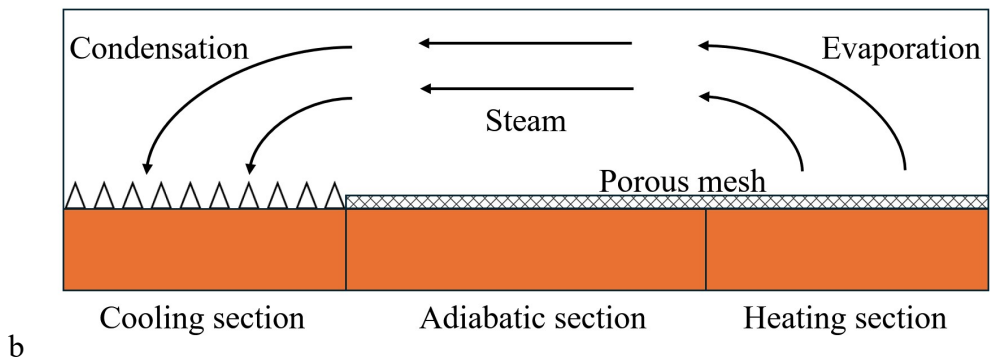

b

c

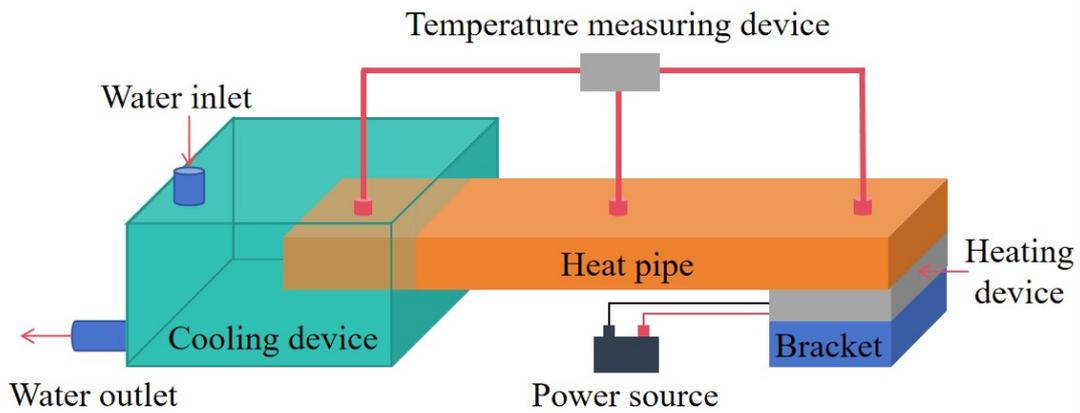

Figure 5  Device using evaporation to draw condensed liquid. **a**, Connecting the right end of the copper plate to the copper foam placed on a heating platform. **b**, In the closed container, the steam from the heating section is directed to the spiked surface for condensation, then the condensed liquid is drawn to the heating section by a porous mesh, forming a heat pipe. **c**, Schematics of the heat pipe experiment. The outer shell of the heat pipe is made of aluminum, with specified dimensions for length 357mm×60mm×20mm, and thickness 1.4mm. Its power reaches 480W.

**The suction effect of condensation**. In the experiment of Figure 1, since the density of steam is less than that of air, some of the steam jet from the outlet will inevitably flow upward into the surrounding air and fail to condense into water. However, the optimal condensation efficiency of a spiked surface reaches 94.1%, which indicates that the surface has a significant suction effect on the steam flow. To confirm this speculation, we conduct the following experiment as shown in Figure 6 and Supplementary Movie 8.

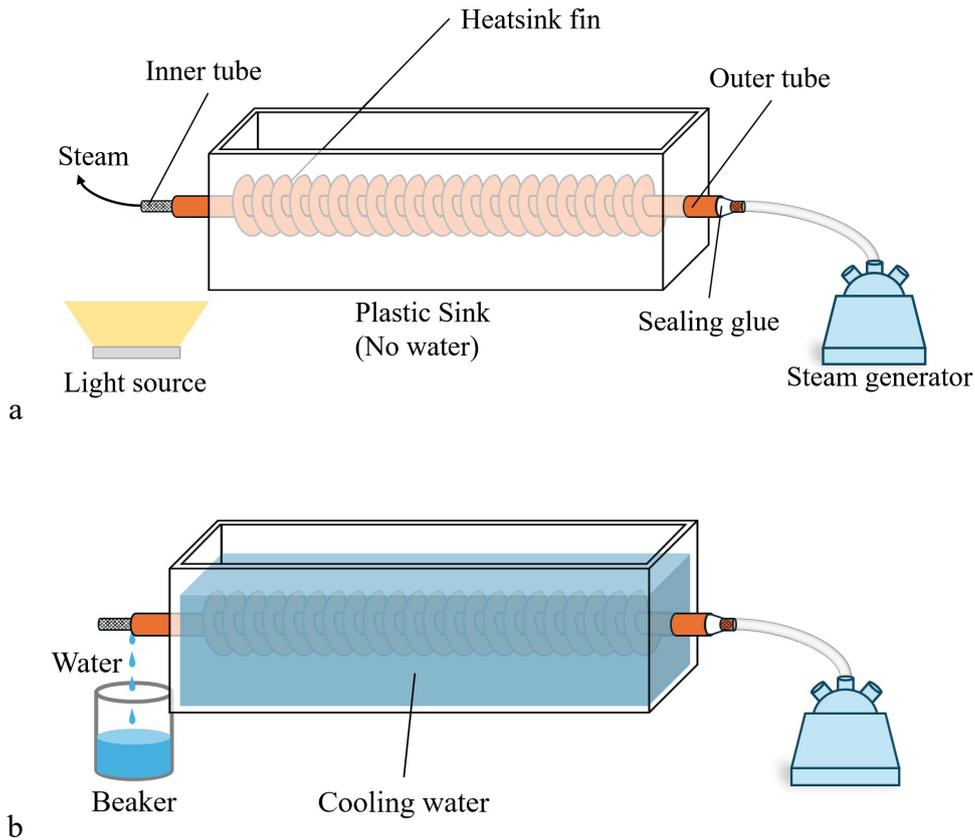

Figure 6  Device for the suction effect of condensation. Two coaxial tubes are placed in a plastic container, with the outer tube being a circular copper tube wrapped with helical fins for heat transfer, and the inner tube being rolled up by a porous mesh. The right end of the inner tube is connected to the steam outlet. **a**, When no cooling water is added to the plastic container, a large amount of steam flows out from the left end of the inner tube. **b**, After adding cooling water (1°C) to the container, the condensed liquid flows out from the bottom of the outer tube on the left end.

We prepare two coaxial tubes and fixed them in a plastic container. The outer cylinder is a round copper tube with spiked inner surface, and its core section is wrapped with helical fins for heat transfer. The left end is open, and at the right end, the annular gap between the inner and outer tubes is sealed with glue. The inner tube is made of porous mesh (180 mesh), with the right section connected to the steam outlet, while at the left end, the steam can flow out freely. The purpose of the porous mesh is to allow the steam to flow from the inner tube to the outer one while providing a certain resistance to this flow.

When no cooling water is added to the plastic container, the outer wall of the copper tube is only in contact with air, resulting in limited heat dissipation. At this time, a large amount of steam flows out from the left end. When encountering colder air, the steam condenses into small water droplets, and the movement of steam flow is displayed under the illumination of LED lights. Thus, we can find that most of the steam flows out from the inner tube, and the amount of steam flowing out from the annular gap between the inner and outer tubes is very small.

Then, we use water (1°C) to cool the copper tube. When the cold water is added to the plastic container, the steam flowing out from the left end decreases significantly and quickly disappears. After a period of time, liquid begins to flow out from the left end of the copper tube, and no more steam is found coming out from the inner tube. This indicates that the high-temperature steam entering from the right end has condensed into liquid and flowed out from the left end of the outer tube. Further experiments suggest that as long as the cooling water is kept below 30°C, the above phenomenon can be maintained; if the temperature of the cooling water is too high, steam will flow out from the left end again, indicating that the steam has not fully condensed inside the copper tube.

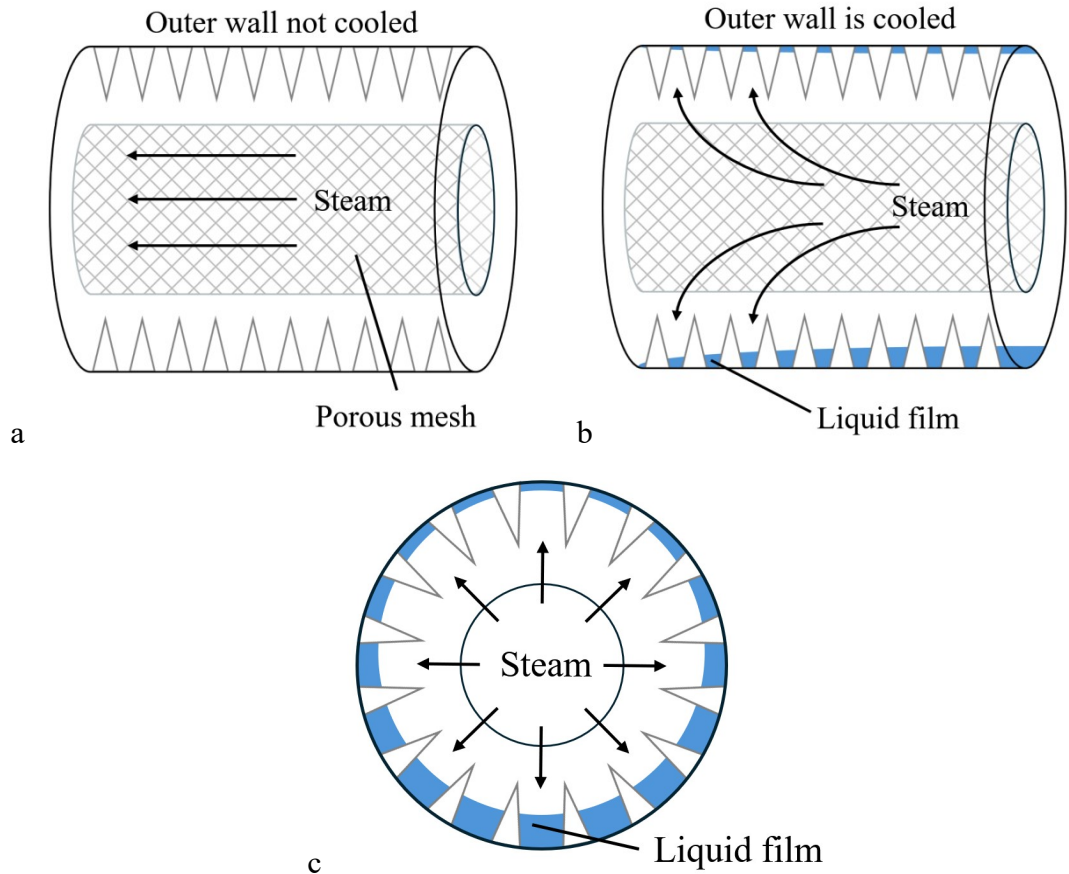

Figure 7 Steam flow and condensation inside the copper tube. **a**, When there is no cooling water in the plastic container, most of the steam flows axially inside the inner tube. **b**, When the outer wall comes into contact with the cooling water, the steam flows through the porous mesh and condenses on the spiked surface. The condensed liquid accumulates at the bottom of the inner wall of the copper tube. **c**, Cross-sectional view during condensation.

Based on the experimental observations, we can infer the situation inside the copper tube: when there is no cooling water in the plastic container, the steam is mainly concentrated in the inner tube and flows axially, with only a small amount of steam passing through the porous mesh and entering the annular gap; when the outer wall of the copper tube is in contact with cooling water, the steam flow can pass through the porous mesh and condense into liquid on the spiked surface, then collect at the bottom of the inner wall, causing the water level of the condensate to continuously rise and eventually overflow from the outlet. If we continuously replace the cooling water in the container to keep the water at a low temperature (<30°C), the process can continue. Therefore, there must be a mechanism that makes the steam flow continuously pass through the porous mesh.

Comparing the directions of the two steam flows in Figure 7, we can infer that the condensation on the solid wall has a significant suction effect on the steam. In heat pipes, although condensation of gas occurs in the cooling section, the traditional theory considers the capillary force as the sole driving force for the system circulation. The driving mechanism generated by phase change has rarely been mentioned in previous literature. The only works we know are given by Liu[36] and Zhu[37], while they did not give clear experimental evidence to prove the existence of the phase change force. Here, the experimental results shown in Figures 6 & 7 demonstrate that the suction force generated by condensation can overcome the resistance of the porous mesh and drive the steam flow towards the condensation side, significantly altering the flow direction. Therefore, for phase-change heat transfer devices, this force can provide a new perspective for understanding the operation mechanism of the system, and is expected to provide a new driving force for self-driven cooling systems through subsequent design to improve condensation efficiency.

In summary, we report a unique method that utilizes superhydrophilic and superhydrophobic coatings as well as millimeter-scale spiked surfaces to achieve highly efficient condensation of steam and timely removal of liquid. We reveal that the combination of a superhydrophobic spiked surface with a superhydrophilic bottom groove offers the best condensation efficiency. This method does not rely on gravity, and therefore can still be applied in microgravity environments, such as in space. The condensation efficiency of the horizontal plate surpasses that of the traditional vertical plate, while the heat flux density reaches $973.8 kW/m^2$. Furthermore, we demonstrate that the condensation on the spiked surface has a significant suction effect on steam. This suction force can be used to drive directional flow of high-temperature steam, promising to provide new circulation power and enhance heat transfer efficiency for phase-change heat transfer devices such as heat pipes. This discovery opens up a new pathway for steam condensation, enriches the knowledge in the fields of fluid dynamics and surface science, and demonstrates its tremendous potential in enhancing heat transfer.

# Method

**Materials**. In this experiment, the copper plates, copper tubes for condensation and copper blocks used for cooling, were all made of red copper purchased from Dongguan Yuexin Metal Material Co., Ltd. The processing of copper plates, copper tubes and copper blocks was completed by Zhuangshi Liquanya Machinery Hardware Factory in Ningbo Zhenhai District using Beijing Andejianqi AR35MA precision CNC electric spark wire cutting machine (machine tool processing accuracy ± 0.001). The parameters of the condensation zone in Figure 1 are displayed in Table 2.

Table 2  The parameters the condensation zone in Figure 1

| $h$ | $L_1$ | $L_2$ | $L_3$ | $L_4$ | $L_5$ | $L_6$ | $L_7$ |
|---|---|---|---|---|---|---|---|
| 2 | 1 | 0.5 | 1 | 0.5 | 0.5 | 1 | 1.5 |
| 3 | 1.5 | 0.5 | 0.54 | 0.5 | 2 | 0.54 | 1 |
| 5 | 1.5 | 0.5 | 0.54 | 0.5 | 4 | 1.61 | 3 |

All values are given in mm.

Spherical condensing tubes were purchased from Jiangsu Ronglipu Scientific Instrument Co., Ltd. Beakers and measuring cylinders (BOMEX, accuracy 0.1ml) were purchased from Beijing Glass Trading Center Co., Ltd. Glass rods were purchased from Changde Bikmam Biotechnology Co., Ltd. Ethanol (concentration 95%), three-neck flasks and rubber stoppers were purchased from Shanghai Yongchuan Biotechnology Co., Ltd. The water used in the experiment was ultrapure water produced by UPT-11-10T manufactured by Youpu. Superhydrophobic XN-207C coating (fluorosilicone modified silica) and superhydrophilic XN-504A coating (alkoxysilane functionalized polymer aqueous dispersion) were purchased from Dongguan Weijing Nanomaterials Co., Ltd. Zeolite (size≤10.0um) was purchased from Bidepharm Co., Ltd.

**Surface treatment.** The spiked surface were machined by a VMC1160 milling machine. Before the spray coating, the copper plate was first placed in an ethanol solution and cleaned for 5 minutes in an ultrasonic cleaning machine (Lichen). After drying, the copper plate was placed in dilute sulfuric acid solution (0.5mol / L) to remove the surface oxides, then rinsed with deionized water for 5 minutes, and finally dried with nitrogen.

For SOGO and SIGI coatings, the corresponding superhydrophobic XN-207C coating and superhydrophilic XN-504A coating were sprayed using a spray gun (the air pump was set to 0.2Mpa, while three passes of the spray gun were performed), and then placed in a vacuum drying oven (TAISITE, China) setting at 150°C for half an hour.

For the SOGI coatings, first, the superhydrophilic coating was coated on the surface of the copper plate. Then a 0.01mm thick polyethylene film was gently covered on the spiked surface, and the film was pressed to the bottom of the spike with a sponge, while the spike punctured the polyethylene film, and three layers of polyethylene film were repeatedly covered. After that, the spiked surface was sprayed with a superhydrophobic coating, dried and then three layers of film were removed. So it can create a superhydrophobic effect on the spiked surface, while the bottom capillary groove surface maintains superhydrophilic characteristics.

For the SIGO coatings, since the superhydrophobic coating will cover the superhydrophilic layer, the process of this combined coatings is different from that before. We first used a syringe to apply the superhydrophobic coating to the microgrooves, then place the copper plate in a 150°C drying oven for 10 minutes, then covered three layers of polyethylene film as described above. After that, we sprayed the superhydrophilic coating on the spiked surface, dried and then removed three layers of film. So the surface of the microgrooves has a superhydrophobic property and the spiked surface is superhydrophilic.

To assess the durability of the coatings, we subject both the superhydrophobic and superhydrophilic surfaces to a two-hour condensation test. Interestingly, the properties of the coatings show no significant changes from before to after the experiment, suggesting their stability over time.

**Observation instruments for condensation zone**. The process of condensation on the spikes was recorded using a high-speed camera (Photron, FASTCAM SA1.1) at a frame rate of 200 frames per second. The surface temperature distribution of the copper plate was recorded using an infrared thermal imaging camera (Telops, FSAT-M100K).

**Wettability characterization**. We characterized the superhydrophobic and superhydrophilic coatings on the surface of the copper plate by measuring the static contact angle using a contact angle measurement instrument (Kino SL250). After placing 0.5 mL of deionized water in the liquid injection device, a droplet of 2 μL was placed on the surface and measured using the sessile drop method. During the titration, the curve obtained by the Young-Laplace equation was used for fitting. By combining mathematical model fitting with the actual contour measurement, the corresponding contact angle was determined. The reading resolution was 0.01° with a testing accuracy of ±0.5°.

**Microstructure characterization**.We obtained the microstructural images of the coatings on the copper plate using a scanning electron microscope (Hitachi SU5000, SEM). The three-dimensional microstructure, surface characteristics, and roughness of the superhydrophobic and superhydrophilic coatings were measured using a laser scanning confocal microscope (Zeiss LSM900).

**Generation of high-temperature steam**. The generation device of high-temperature steam is mainly composed of three parts: an intelligent electric heating jacket (LICHEN ZNHW, 2000mL) for heating the liquid, a three-neck flask (2000mL), and a silicone tube (1m long, outer diameter 25mm, inner diameter 19mm), as shown in Figure 1. The heating jacket is set to 160°C to ensure that the heating device is at maximum power. The three-neck flask is placed in the heating jacket, with rubber plugs blocking the two side ports, and a temperature sensor is inserted in one port. After wrapping a rubber insulated tube (thickness 7mm) on the outer wall of the silicone tube, one end is inserted into the middle port of the three-neck flask. The steam flows out from the outlet and condenses on the copper plate. Before heating, we put 1200mL of deionized water in the flask and add 5g of zeolite powder with a particle size less than 10μm to prevent bumping.

**Cooling device**. The cooling device is mainly composed of a copper block with dimensions of 80mm length, 40mm width, and 6mm height, and a constant-temperature circulation tank (huber, cc-k60), as shown in Figure 1. The silicone tube wrapping a rubber insulated tube connects these two parts. For the circulation tank, the working temperature is set to 5°C, the rotation speed of the pump is set to 2500rpm, and an appropriate amount of water is placed for cooling.

**Data availability**

The authors declare that the data supporting the findings of this study are available within the paper and its supplementary information files. Source data are available upon reasonable request.

**Declaration of Interests.** The authors report no conflict of interest.

**Author contributions.** Kai-xin Hu made substantial contributions to the conception of the work, wrote the paper for important intellectual content and approved the final version to be published. He is accountable for all aspects of the work in ensuring that questions related to the accuracy or integrity of any part of the work are appropriately investigated and resolved. Yin-Jiang Chen conducted steam condensation experiments, prepared the materials and drew the figures. Bin-Wei Tang conducted suction effect experiments. Yue-Qun Tao and Qiu-Sheng Liu provided suggestions for the experimental scheme. Qi-shen Chen provided editing and writing assistance.

**Funding.** This work has been supported by the National Natural Science Foundation of China (No.12372247), Ningbo Municipality Key Research and Development Program (No. 2022Z213) and the China Manned Space Engineering Application Program—China Space Station Experiment Project (No. TGMTYY14019).